\documentstyle[preprint,eqsecnum,aps]{revtex}

\begin{document}
\draft
\preprint{}
\title{MAGNETIC AND CHARGE FLUCTUATIONS
IN HIGH-$T_c$ SUPERCONDUCTORS
\\}
\author{H. A. Mook,$^1$ F. Do$\rm\breve{g}$an,$^2$ and B. C. Chakoumakos$^1$}
\address{
$^1$Oak Ridge National Laboratory, Oak Ridge, 
Tennessee 37831-6393\\
$^2$Department of Materials Science and Engineering,
University of Washington, Seattle, Washington 98195\\
}
\maketitle
\begin{abstract}
Neutron scattering has been used to study the spin fluctuations in 
the YBa$_2$Cu$_3$O$_{7-x}$ and Bi$_2$Sr$_2$CaCu$_2$O$_8$ materials. 
Evidence is found for both incommensurate fluctuations and a commensurate 
resonance excitation. Measurements on the lattice dynamics 
for YBa$_2$Cu$_3$O$_{6.6}$ show incommensurate structure that 
appears to stem from charge fluctuations that are associated with the spin fluctuations. 
\end{abstract}
\pacs{PACS numbers: 74.72.Bk, 61.12.Ex}

\narrowtext
\section{INTRODUCTION}
Neutron scattering measurements continue to provide information of 
direct relevance to some of the most important issues in the high-$T_c$ cuprate superconductors. 
The magnetic excitations of these materials are the spin fluctuations, and recent 
measurements have shown that the low-energy spin fluctuations in YBa$_2$Cu$_3$O$_{6.6}$ 
(YBCO$_{6.6}$) are incommensurate in nature while a commensurate excitation 
that is relatively sharp in energy called a resonance is found at about 35 meV \cite{daimook}.  
The incommensurability was originally discovered by the filter integration technique \cite{mook1} 
that integrates over the outgoing neutron energy in a direction along $c^\ast$ and 
thus provides a high data collection rate for the study of lower dimensional excitations.  
The disadvantage of the technique is that no discrete energy information is available.  
Thus when a discovery is made by the integration technique further measurements are made by 
triple-axis or time-of-flight techniques to determine the energy spectrum.  
Figure 1a shows the direction of the integrating scan that is made through the point 
$(\pi,\pi)$ to observe the incommensurate fluctuations shown by the dots at the 
position $\delta /2$ from the commensurate position.  Such a scan employs high 
resolution along the scan direction, but coarse resolution perpendicular to the scan 
direction and thus cannot determine the exact wave vector position of the incommensurate peaks.  
The result of the scan for YBCO$_{6.6}$ is shown in Fig. 1b.  The $(\pi,\pi)$ position is 
at (0.5, 0.5) in reciprocal lattice units (r.l.u.). 

A recent measurement using a pulsed spallation source with time-of-flight 
energy determination has used a  two-dimensional position-sensitive detector 
bank to determine the wave vector and energy of the YBCO$_{6.6}$ incommensurate peaks \cite{mook2}.  
The peaks are found to be along the $(0, \pi)$
and $(\pi, 0)$ directions as shown in Fig. 1a.  
The peaks are therefore not exactly on the scan direction shown in Fig. 1a, but are above 
and below it being observed in the integrated scan through the relaxed vertical resolution.  
The peaks in the scan in Fig. 1b are found at about 0.055 on either side of the commensurate point. 
The wave vector of the incommensurate scattering is then $\sqrt 2$ 
times 0.055 from the geometry in Fig. 1 and this value must be multiplied by $\sqrt 2$ 
again as the scan is in units of $(h,h)$ giving $\delta /2=0.11$ or $\delta=0.22$. 
The number determined from the time-of-flight measurement is $0.21\pm0.02$ r.l.u..  
This incommensurability is essentially identical to that observed in similarly doped \cite{yamada}
La$_{1-x}$Sr$_x$CuO$_4$ (214).  It was found that the intensities and the correlation 
lengths are also very similar in the (214) single-layer and YBCO (123) bilayer materials, 
thus the low energy spin fluctuations appear to be universal for the high-Tc cuprate materials 
measured to date.

\section{MAGNETIC FLUCTUATIONS IN BSSCO}
Bi$_2$Sr$_2$CaCu$_2$O$_8$ (BSSCO) or (2212) is a high temperature superconductor 
of considerable importance as one can cleave the material easily to obtain a good surface 
for photoemission measurements and other surface sensitive techniques.  Unfortunately it is 
difficult to obtain the large single crystals needed for neutron scattering for the BSSCO composition.  
We were successful in growing rods of the material which have a $[110]$ reciprocal lattice direction 
along the rod direction (using the same definition of $a$ and $b$ as for YBCO).  It was then 
possible to align a number of these rods together to form a sample of 35 grams for neutron studies.  
We have a crystal with one set of the [110] directions aligned and this permits some information 
to be obtained.

A difficulty with the sample for the study of magnetic excitations is that the material 
has  bilayers similar to those in YBCO and thus we expect that the magnetic fluctuations are 
coupled in a similar way.  In this case the low energy magnetic fluctuations have no 
structure factor unless a finite value of $c^\ast$ is used and $c^\ast$ 
is randomly orientated perpendicular to the aligned direction of the crystal.  
One must then set the spectrometer to sample a point off the [110] direction 
which means that only some of the $c^\ast$ values desired fall within the resolution 
volume of the spectrometer, considerably reducing the magnetic signal. 
	
One might expect that a resonance excitation might exist in BSSCO in a 
similar way as for YBCO and one can search for it in the same way \cite{mook3} 
using a triple-axis spectrometer.  Our sample of BSSCO has an oxygen composition 
near optimum doping and we expect that the resonance will not be observable much 
above $T_c$ in this case.  The simplest experiment is thus to scan energy at the momentum 
value where the resonance is expected and take the difference between data taken well 
below $T_c$ and data taken above $T_c$. This works in YBCO, but fails in BSCCO for two related reasons.  
The first is that the signal is small relative to the phonons as we cannot achieve the full 
magnetic intensity that would be available at a fixed value for $c^\ast$.  The second problem is 
the phonons are much more temperature sensitive in BSSCO than they are in YBCO so that the 
difference in data at high and low temperatures strongly reflects the phonon differences.  
The only way to circumvent these difficulties is to use a polarized neutron beam to 
isolate the magnetic scattering.  This unfortunately results in even less intensity 
as polarized beams are much weaker than unpolarized ones.  Nevertheless, after 
considerable counting reasonable results were obtained.  Momentum values near 
(0.5, 0.5) and (1.5, 1.5,) were both tried and similar patterns were obtained with the 
magnetic scattering near (1.5, 1.5) being considerably weaker because of the magnetic form factor.  
The  results are shown in Fig. 2.  A peak in energy about 10 meV wide, which is equal to the 
energy resolution of the experiment, is observed at 10 K while the result at 100 K appears 
rather featureless.  The peak is only found in the spin flip channel guaranteeing that it is magnetic.  
The peak is observed at about 37 meV which is near the value expected if the $T_c$ of the BSSCO of 
84K is scaled to that of YBCO for the same doping level.  The results strongly 
suggest that BSSCO has a resonance excitation rather similar to that of YBCO.  
However, the results should be checked with single crystals when they become available.  

The same BSSCO sample was employed to search for magnetic incommensurate fluctuations. 
The integration technique was used in the same way as for YBCO$_{6.6}$. The experiment works 
in a similar manner except that the integration now takes place over the directions perpendicular 
to the [1,1,0] direction and thus is only partly along $c^\ast$.  For two-dimensional 
scattering from bilayers this results in an intensity loss in the magnetic signal.  
However, we see from Fig. 1 the magnetic signal in the integration technique is substantial 
so that an intensity loss may be tolerated.  The results of the measurement are shown in 
Fig. 3.  Fig. 3a shows data presented in the same way as for the YBCO$_{6.6}$ in Fig 1.  
The results suggest the possibility of small incommensurate peaks although the counting 
errors are larger than desired.  The data shown are from a number of runs averaged together.  
Fig. b-d show one of the satellite peaks measured at different temperatures.  
In this case the background was obtained by a 30 deg. rotation of the sample 
relative to the position where the scan is performed.  The magnetic signal is expected to 
be small in the 30 deg rotation case which samples reciprocal space well removed from (0.5, 0.5).  
The signal decreases with temperature as would be expected for a magnetic excitation.  The peak is 
broad so the center is hard to determine accurately with the errors involved, however, the 
peak appears to be centered at about 0.42 r.l.u. or 0.08  r.l.u. units from the (0.5, 0.5) position.  
If the magnetic satellites are arranged as in Fig. 1, $\delta$ would be about 0.32.  
The value of $\delta$ for fully doped 214 materials is about 0.25 so the $\delta$ value for 
BSSCO appears to be somewhat larger than for the 214 materials assuming the same type of 
incommensurability.  However, the BSCCO measurement has sizable counting errors 
and we do not know the actual pattern of the incommensurability.  A measurement on a 
single crystal is needed for a more accurate determination.  Until such an experiment 
is performed the present result suggests that BSSCO has low energy incommensurate fluctuations 
which may be similar to those found in 123 and 214 materials. 

\section{CHARGE FLUCTUATIONS IN YBCO}  
Neutrons cannot measure charge directly, but can observe a change in the mass density 
that is either static or dynamic.  One can assume that static mass displacements reflect 
static charge ordering and such effects have been observed in the 214 cuprate materials in special 
cases \cite{tranquada}.  Dynamic charge ordering has not been observed so far and the present results 
serve as an indication that this occurs in the YBCO materials.  
We have made measurements on the same YBCO$_{6.6}$ sample in which the 
incommensurate magnetic fluctuations are observed.  Again we start with the 
integration technique except we examine the region around the peaks of the 
reciprocal lattice stemming from the atoms rather than from the magnetism.  
Fig. 4 shows scans around the (1, 0) reciprocal lattice peak.  Data are again shown 
that use a measurement at 300K as a background.  As the sample is cooled distinct 
peaks form on both sides of the (1, 0) peak that we assume reflects a dynamic 
incommensurate mass fluctuation that can be considered to stem from an incommensurate 
charge fluctuation.  However, we have not completely ruled out magnetic effects.  
We note the peaks are small, being an order of magnitude smaller than the magnetic 
satellite peaks shown in Fig. 1. The scan is along the (h, 0) reciprocal lattice or the $(\pi, 0)$ 
direction and thus is along the direction of the magnetic incommensurate scattering. 
The charge fluctuation peaks are about 0.22 in r.l.u. units from the commensurate 
position so that the $\delta /2$ value for them is 0.22 or twice the wave vector of the 
incommensurate magnetic satellites.  However, the absolute direction of the 
incommensurate wave vector cannot be determined with the integration technique 
and the peaks could be at a wave vector off of the $(\pi, 0)$ direction.  
Fig. 4d shows an identical measurement for YBCO$_{6.35}$ 
which has a much lower oxygen content. No peaks are visible at 0.22 r.l.u. for this material. 

Work has been underway with triple-axis spectrometry to 
determine the energy spectra of the charge fluctuations, but that 
work is still incomplete.  It has been noted however, that certain phonon 
branches show anomalies at the wave vector of the charge fluctuations.  
The origin of the charge fluctuations is not clear.  It would seem extremely 
likely that the magnetic and charge fluctuations stem from the same source.  
Obviously the observation of charge fluctuations strongly suggest a dynamic 
striped phase in YBCO$_{6.6}$.  However, other possibilities exist including Fermi 
surface effects or dynamic charge density waves.  The next step is to determine the 
energy spectra and absolute wave vector of the incommensurate charge scattering.

\section{CONCLUSION}
We have shown new neutron scattering results for the cuprate superconductors.  
Measurements on a BSCCO sample of crystals with a [110] direction aligned show 
strong evidence for a resonance excitation and indications of incommensurate magnetic 
fluctuations.  It would be good to have these results confirmed by a high quality 
single crystal.  For YBCO$_{6.6}$ clear dynamic incommensurate peaks are observed 
at low temperatues on either side of the (1, 0) reciprocal lattice peak.  
Since these are found at positions relative to the crystal reciprocal lattice they are 
assumed to stem from mass fluctuations driven by charge fluctuations.  
No such peaks are found for a YBCO$_{6.35}$ sample.  The wave vector of the 
charge fluctuation peaks is twice that of the magnetic fluctuations 
if we assume the charge peaks are on the $(0,\pi)$ direction.  
It would seem likely the magnetic and charge excitations are related.  
The results give support to a dynamic striped phase model for the cuprate superconductors.      

This research was supported by the US DOE under 
Contract No. DE-AC05-96OR22464 with Lockheed Martin 
Energy Research, Inc.

\begin{figure}
\caption{(a) shows the position of the magnetic incommensurate peaks 
found for YBCO$_{6.6}$.  The direction of the energy integrating 
scan used to discover the incommensurability is shown by the arrow.  (b) 
shows the result of the integrating scan.  The measurement was made at 10K 
while 300K data is used as a background since little magnetic scattering is found at 300K.  
}
\label{autonum}
\end{figure}

\begin{figure}
\caption{(a) Polarized neutron measurement of the resonance excitation in BSCCO.  
(a) shows the result at 10K while (b) is the 100K result. The experimental resolution is 
about 10 meV so that the peak in (a) is resolution limited. A background 
determined by moving the analyzer off the Bragg position has been subtracted. 
A number of runs were averaged to obtain the counting errors shown. 
}
\end{figure}

\begin{figure}
\caption{Integrating scans made to search for incommensurate magnetic 
fluctuations in BSCCO.  (a) shows a scan made in the same way as the scan for YBCO$_{6.6}$ 
shown in Fig. 1(b).  b-d show scans made over the lower wave vector incommensurate position.  
The scans are made with more momentum steps and several scans are averaged to obtain 
lower counting statistics.  The background used is obtained by a rotation of the 
sample so the background samples a region removed from the (0.5, 0.5) magnetic position.}
\end{figure}

\begin{figure}
\caption{Intergrating scans made around the crystal reciprocal lattice position 
(1, 0) for YBCO. (a)Ð(c) show that incommensurate structure appears at 
low temperatures at a wave vector twice that of the magnetic incommensurate satellites for YBCO$_{6.6}$.  
A measurement made in the same way for YBCO6.35 is shown in (d).}
\end{figure}

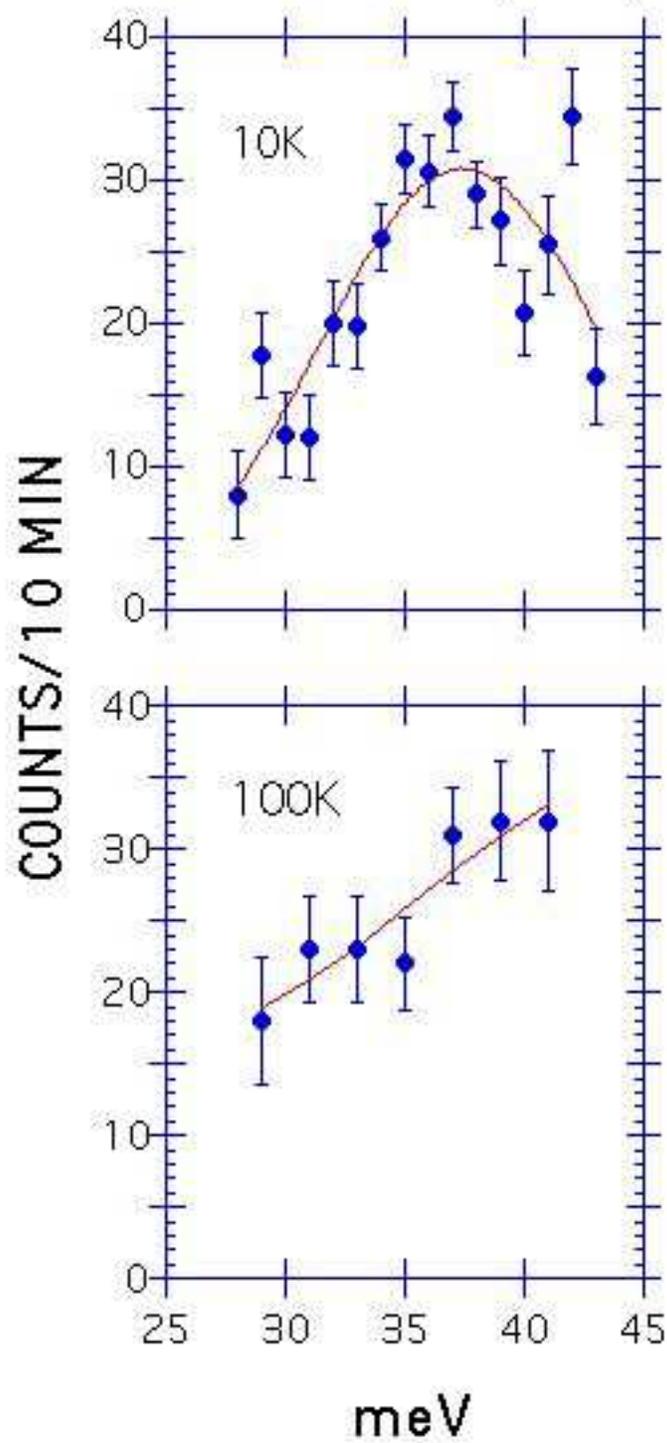
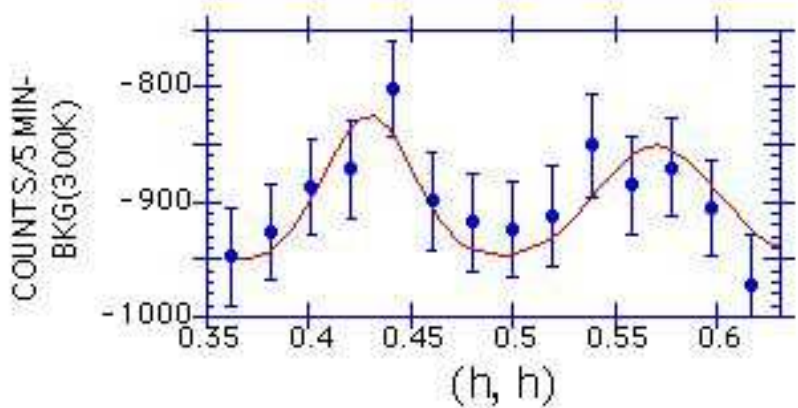
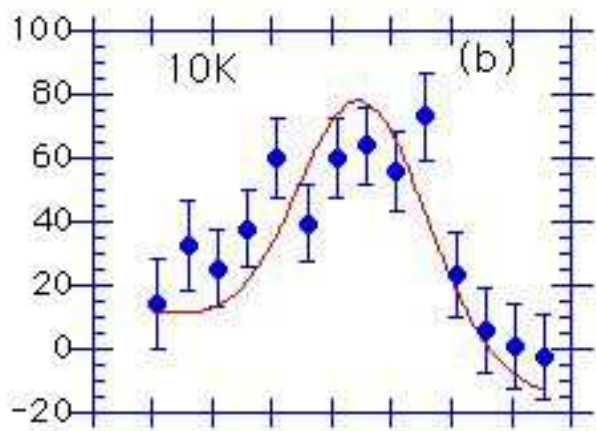
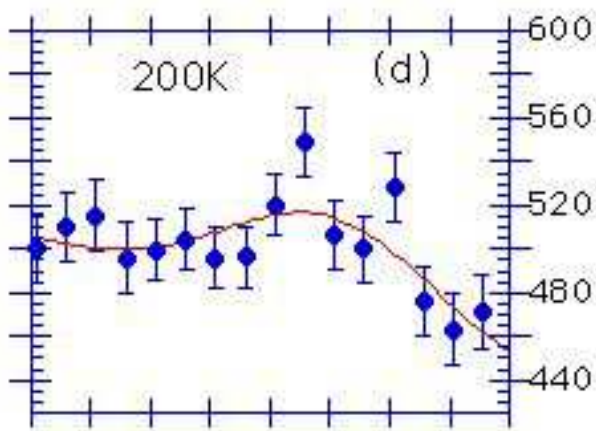
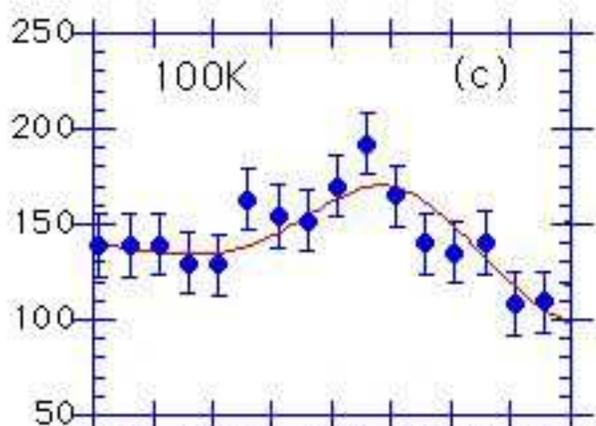
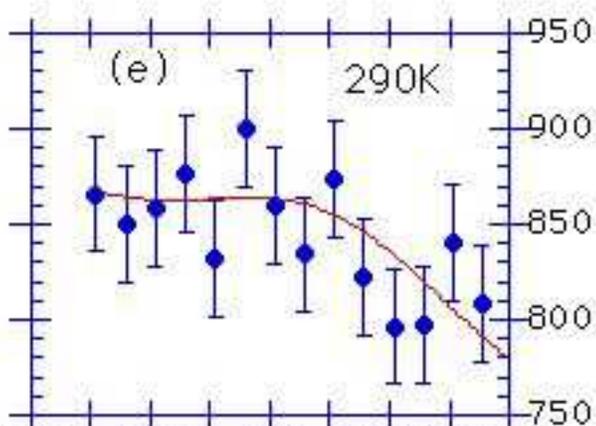
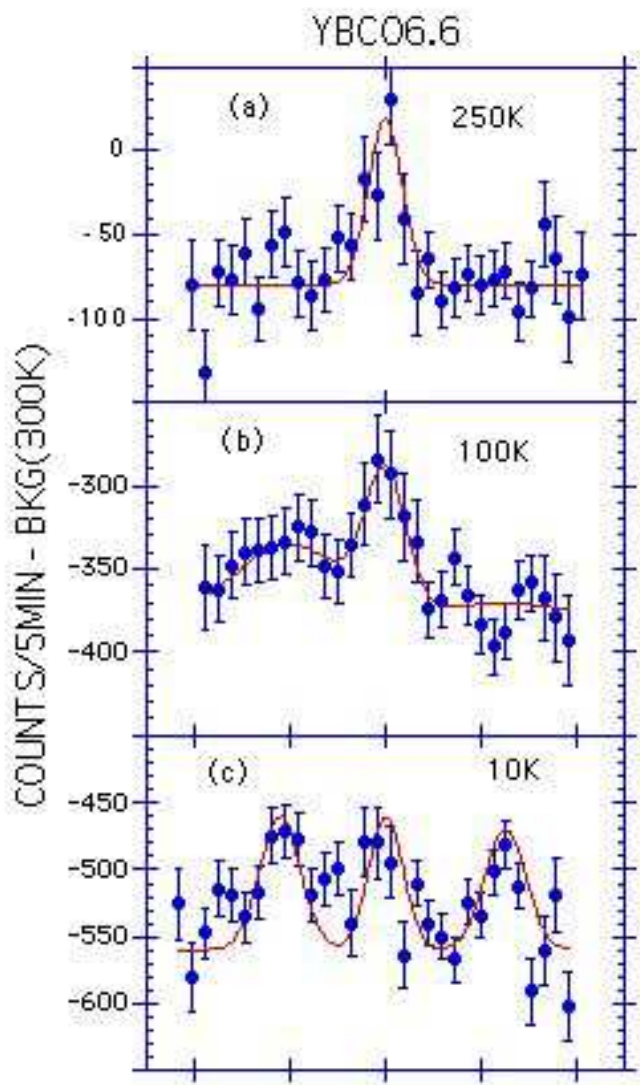
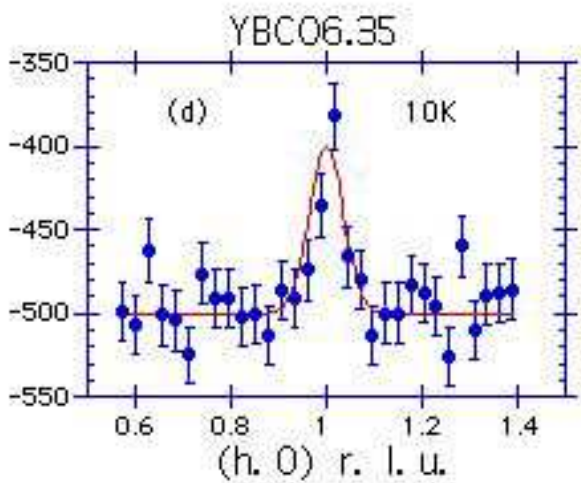
\end{document}